# Aesthetics of Connectivity: Envisioning Empowerment Through Smart Clothing

Yannick Kibolwe Mulundule[1, *], Cheng Yao[1], Amir Ubed[1], Abdiaziz Omar Hassan[1]

1. Zhejiang University Ningbo Innovation Center, China

**Abstract**— Empowerment in smart clothing, which incorporates advanced technologies, requires the integration of scientific and technological expertise with artistic and design principles. Little research has focused on this unique and innovative field of design until now, and that is about to change. The concept of 'wearables' cut across several fields. A global 'language' that permits both free-form creativity and a methodical design approach is required. Smart clothing designers often seek guidance in their research since it may be difficult to prioritize and understand issues like as usability, production, style, consumer culture, reuse, and end-user needs. Researchers in this research made sure that their design tool was presented in a manner that practitioners from many walks of life could understand. The 'critical route' is a useful tool for smart technology implementation design, study, and development since it helps to clarify the path that must be taken.

## Introduction

Smart clothing combines aesthetics and cultural appropriateness with electronics and computers to increase its functionality. Fashion and textile design must be merged with science and technology to create appealing smart garments with advanced technologies. This research examines a new, understudied design field. Beyond fashion, smart clothing development creates problems. Effective apparel requires creative and thorough design. Instead of describing its purpose and value to the end user, fashion design might be contextualized using art, literature, or music [1]. This study proposes a comprehensive framework for designing smart clothing that balances aesthetics and functionality, enabling designers, engineers, and researchers to develop cutting-edge and practical smart garments.

## Background

Smart clothing with embedded technology is becoming increasingly widespread in recent years, with numerous applications in sports, fitness, workplace, healthcare, and elderly products. Before garment design, end-user needs must be considered. Visual, product and industrial design may help identify these needs. In recent decades, competitive, extreme, and leisure sports and health and fitness have increased, driving functional gear sales. Fiber and fabric technology and new garment production methods have made performance gear more attractive and detailed. These developments began in the military. Marketing hype has affected extreme sports, adventure travel, work dress, and health clothing. Sportswear is comfortable and low maintenance for the growing elderly population. Modern "smart" technologies improve clothing comfort, style, and function.

The combination of smart fabrics and design for users offers a unique opportunity to push the boundaries of smart garments. The purpose of this paper is to examine how exactly the integration of such elements into smart garments will help to enhance their effectiveness and attractiveness addressing specific issues such as comfort, environmental adaptability, and aesthetic design [1]. This paper is designed for an interdisciplinary audience including designers, technologists, and HCI researchers focused on the integration of aesthetics and functionality within smart clothing. It explores how these fields converge to enhance user empowerment through innovative wearables. The theoretical framework synthesizes insights from design theory, human-computer interaction, and user-centric technology development. This framework posits that successful smart clothing must balance aesthetic appeal with technological efficacy and user-centric design to truly empower users. However, the lack of a systematic methodology for designing smart clothing has hindered the development of innovative and user-friendly garments. The proposed framework aims to fill this gap by providing a structured approach to integrating technology into clothing.

- A. *Research Question*
  How can the integration of aesthetics and technology in smart clothing enhance user empowerment and functionality while addressing specific user needs and cultural considerations?

- B. *Targeted Audience*
  This paper is designed for an interdisciplinary audience including designers, technologists, and HCI researchers focused on the integration of aesthetics and functionality within smart clothing. It explores how these fields converge to enhance user empowerment through innovative wearables.

# Methodology

The proposed framework is based on a mixed-methods approach, combining both quantitative data collection and analysis. A literature review was conducted to identify key factors influencing the design of smart clothing, including material science, textile engineering, and human- computer interaction. This research uses secondary data collection to explore the aesthetics of connectivity in smart clothing. It searches for articles discussing technology incorporation, design, and user empowerment. The sources are categorized based on design, usability, technology, and market trends. User experience research on comfort, functionality, and cultural aspects is also included. The review uses peer-reviewed scientific journals, databases, and online resources. Market trend analyses provide insights into consumer behavior and the future of smart garments as a market product.

# Keyword Selection

The literature search focused on smart clothing, user empowerment, and technology-aesthetic integration. Key terms included "smart clothing," "wearable technology," "aesthetic design," "user control," and "smart surroundings" to explore technology and design principles, user experience, and peripheral areas influencing wearables.

# Literature Review Process

The literature review process was structured to ensure a thorough and methodical examination of existing research. The following steps outline approach was undertaken

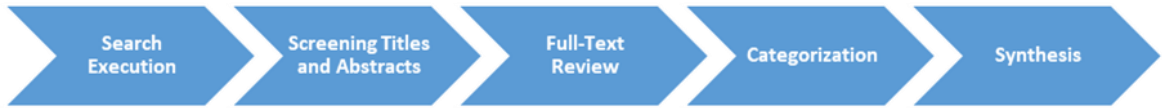

## Related Works

Wearable technologies: Several studies have explored the integration of sensors, conductive materials, and microchips into garments that enable them to collect and transmit data, interact with devices, and provide real-time feedback to the wearer to enhance their experience. These works have focused on different aspects of wearables, including aesthetics, textiles, texture, materials, and functionality. For example, smart clothing can monitor vital signs, track physical activity, or even provide haptic feedback to the wearer. Understanding the current state-of-the-art in smart clothing within wearable technologies will provide a foundation for this study.

Deepti, Hoang and Bernd (2021), mentioned that, to this end, smart clothing should support the following six key factors: Usability, Functionality, Safety, Durability, Comfort and Fashion [2]. On the other hand, Gemperle and Colleagues described thirteen guidelines that should be considered while designing wearable technology [3].

## Proposed Framework

Although there has been considerable research on incorporating technology into clothing projects in the last ten years, there is a noticeable lack of comprehensive frameworks for designing smart clothing. This has led to a shortage of standardized guidelines and best practices in the field. This study proposes a comprehensive framework for designing smart clothing. The suggested framework aims to offer a systematic methodology for designers, engineers, and researchers to develop cutting-edge and practical smart garments that meet user requirements.

Table 1

Design framework

| Design pillar | Design element |
|---|---|
| Technology user relationship | User-centered design, Easy to use interfaces, Feedback mechanisms, Personalization, Safety and security |
| Technology environment relationship | Environmental factors, Power consumption, Connectivity, E-waste reduction |
| User environment relationship | Comfort and wearability, Contextual awareness, Cultural sensitivity, inclusivity |
| Interdisciplinary collaboration | Collaborative design process, Iterative deign approach |
| Conceptualization | User research, Ideas generation, Ideas filtering |
| Design development | Sketching, Computer aided design Prototyping |
| Material selection | Material research, Material combination, Material testing |
| Aesthetics and fashion appeal | Design for style, comfort and wearability |
| Integration and testing | Combination, User test |

## Discussion

The study suggests that aesthetics play a crucial role in determining the effectiveness of smart clothing. These ideas and the dynamics between technical advancement, design variables, and potential user control are discussed in this section. First, the literature emphasizes user centrality in smart garment design. Smart clothing design should prioritize users, expectations, and practicalities, according to [4]. [5] emphasizes product comfort, functionality, and attractiveness daily. Previous study on smart clothing teaches customers that textiles should be practical and comfortable to wear over extended periods. This is consistent with wearable industry trends that show a growing desire for portable, easy-to-use technologies.

Advances in fiber technology and textile engineering can help integrate technology into smart garment design. Conductive textiles and flexible circuit boards enable smart clothing that captures physiological signals, provides real-time feedback, and interacts with the wearer. These technological advances enhance functionality and personalized experiences, while also considering aesthetics and market approval criteria. Eye appeal can boost customer acceptance and usage of smart clothing. The study explores the cultural fit of smart apparel, highlighting the importance of understanding different civilizations' views on clothing. Designers must consider cultural factors when choosing colors, styles, and functions, aiming to boost user acceptability and empowerment by meeting social needs. [6] also shows that smart clothing design and production need sustainable methods. The public's growing awareness of ecological issues has led to a surge in conservancy in fashion, with smart apparel design and manufacturing attracting environmentally conscious buyers. These include using reused, ethically sourced textiles, stitching smartly, and considering garment disposal. Sustainable smart clothing may improve consumers' lives by encouraging them to choose sustainable apparel. The study demonstrates the importance of a holistic approach in wearable technology, integrating design, technology, and user empowerment. It uses the 'critical route' design tool to demonstrate how smart garments can improve user experience through aesthetic and functional integration. The research bridges gaps between Human-Computer Interaction (HCI), textile design, and technology development. HCI researchers gain insights into user interaction with technologically enhanced garments, while textile designers explore new materials without compromising style or comfort. Technologists benefit from integrating advancements into wearable designs, fostering a collaborative environment for innovative solutions in smart clothing development.

## Conclusion

Finally, a design brief with end-user objectives is crucial for developing smart clothes. Designers must be informed about emerging technology, cultural trends, and customer requirements. The clothing industry defines "smart" as abrasion resistance, insulation, and more. Modern technologies impact sports, healthcare, industrial operations, global communications, and fibers. Sustainability requires durable, non-toxic components in recyclable materials and renewable fibers. Designers may collaborate with computer scientists, digital media professionals, biologists, textile technologists, garment engineers, and electronics. A critical route method-based design tool is proposed but may need refinement and validation.

## Acknowledgements

This research was supported by the Fundamental Research Funds for the Central Universities (Grant No. 226-2024-00164), "Leading Goose" R&D Program of Zhejiang (Project No.2023C01216), Research Center of Computer Aided Product Innovation Design, Ministry of Education. We also acknowledge and value the significant contributions made by all individuals who have aided us along the way.